\documentclass[%
 reprint,
superscriptaddress,
 amsmath,amssymb,
 aps,prl
]{revtex4-1}

\usepackage{graphicx}
\usepackage{dcolumn}
\usepackage{bm}

\usepackage[usenames]{color}

\usepackage{pgfplots}
\usepgfplotslibrary{external}
\tikzsetexternalprefix{Figures/Tikz/}

\usetikzlibrary{arrows,shapes,backgrounds,calc,
    positioning,
    intersections}

\newcommand{\kB}{k_{\mathrm{B}}}
\newcommand{\ti}{t_{\mathrm{i}}}
\newcommand{\tf}{t_{\mathrm{f}}}
\newcommand{\Ueff}{U_{\mathrm{eff}}}
\newcommand{\Ereff}{E_{\mathrm{r}}^{\mathrm{eff}}}
\newcommand{\Er}{E_{\mathrm{r}}}
\newcommand{\VL}{V_{\mathrm{L}}}
\newcommand{\VB}{V_{\mathrm{B}}}
\newcommand{\Heff}{H_{\mathrm{eff}}}
\newcommand{\Veff}{V_{\mathrm{eff}}}
\newcommand{\deff}{d_{\mathrm{eff}}}
\newcommand{\dd}{\mathrm{d}}
\newcommand{\tramp}{t_{\mathrm{ramp}}}
\newcommand{\vlatt}{v_{\mathrm{latt}}}
\newcommand{\rr}{\mathbf{r}}

\newcommand{\ket}[1]{\left| #1 \right>} 

\def\be{\begin{equation}}
\def\ee{\end{equation}}

\def\bs#1{\boldsymbol{#1}}

\definecolor{Nathanblue}{rgb}{0.,0.24,0.51}

\tikzset{every axis plot/.style=thick}

\definecolor{Green}{cmyk}{0.9,0.2,1,.0}
\definecolor{Blue}{cmyk}{1,0.4,0,.0}
\definecolor{Red}{cmyk}{0,0.8,1,.0}
\definecolor{Black}{cmyk}{0,0,0,0.9}
\pgfplotscreateplotcyclelist{list2}{%
{color=Red},
{color=Blue},
{color=Green},
{color=Black},
{color=Black},
{color=Green},
{color=Blue},
{color=Red},
{color=Red},
{color=Blue},
{color=Green},
{color=Black},
{color=Black},
{color=Green},
{color=Blue},
{color=Red},
{color=Red},
{color=Blue},
{color=Green},
{color=Black},
{color=Black},
{color=Green},
{color=Blue},
{color=Red},
}

\pgfplotscreateplotcyclelist{list3}{%
{color=Red},
{color=Black},
}

\usepackage{pdfpages} 

\begin{document}

\preprint{APS/123-QED}

\title{Dynamic optical lattices of sub-wavelength spacing for ultracold atoms}

\author{Sylvain Nascimbene}\email{sylvain.nascimbene@lkb.ens.fr}
\affiliation{%
Laboratoire Kastler Brossel, Coll\`ege de
France, ENS-PSL Research University, CNRS, UPMC-Sorbonne Universit\'es, 11 place Marcelin Berthelot, 75005 Paris, France\\
}%
 
\author{Nathan Goldman}%
\affiliation{%
Laboratoire Kastler Brossel, Coll\`ege de
France, ENS-PSL Research University, CNRS, UPMC-Sorbonne Universit\'es, 11 place Marcelin Berthelot, 75005 Paris, France\\
}%
\affiliation{%
CENOLI, Facult\'e des Sciences, Universit\'e Libre de Bruxelles (U.L.B.), B-1050 Brussels, Belgium\\
}%
\author{Nigel R. Cooper}%
\affiliation{%
 T.C.M. Group, Cavendish Laboratory, J.J. Thomson Avenue, Cambridge CB3 0HE, United Kingdom\\
}%
\author{Jean Dalibard}%
\affiliation{%
Laboratoire Kastler Brossel, Coll\`ege de
France, ENS-PSL Research University, CNRS, UPMC-Sorbonne Universit\'es, 11 place Marcelin Berthelot, 75005 Paris, France\\
}%

\date{\today}

\begin{abstract}
We  propose a scheme to realize lattice potentials of sub-wavelength spacing for ultracold atoms. It is based on spin-dependent optical lattices with a time-periodic modulation. We show that the atomic motion is well described by the combined action of an effective, time-independent, lattice of small spacing, together with a micro-motion associated with the  time-modulation. A numerical simulation shows that an atomic gas can be adiabatically loaded into the effective lattice ground state, for timescales comparable to the ones required for adiabatic loading of standard optical lattices. We generalize our scheme to a two-dimensional geometry, leading to Bloch bands with  non-zero Chern numbers. The realization of lattices of sub-wavelength spacing allows for the enhancement of energy scales, which could facilitate the achievment of strongly-correlated (topological) states.
\end{abstract}

\pacs{Valid PACS appear here}
\maketitle

Optical lattices have allowed experiments on ultracold atomic gases to investigate a large range of lattice models of quantum many-body physics \cite{lewenstein2007ultracold}. Their development led to the realization of strongly-correlated states of matter, such as bosonic and fermionic Mott insulators, and  low-dimensional gases \cite{bloch2008many}. 
In its simplest form, an optical lattice consists of  the optical dipole potential associated with a standing wave of retro-reflected laser light. It can be described as a periodic potential $V(x)=U_0\,\cos^2(kx)$, of spatial period $d=\lambda/2$, where $\lambda$ is the laser wavelength and $k=2\pi/\lambda$. More complex optical lattices, such as superlattices \cite{sebby2006lattice,folling2007direct} or two-dimensional honeycomb lattices \cite{soltan2011multi,tarruell2012creating}, can be generated with suitable laser configurations. 
The recoil energy $\Er=h^2/(8md^2)$, where $h$ is Planck's constant and $m$ is the atom mass, sets the natural energy scale for elementary processes, such as atom tunneling between neighboring lattice sites, as well as the temperature range $T\lesssim \Er/\kB\sim100\,$nK, typically required for quantum degeneracy. 

For a large class of models, the physical behavior is dictated by processes associated with even much smaller energies, such as super-exchange or magnetic dipole interactions \cite{lewenstein2007ultracold}. The associated temperature scales remain out of reach in current experiments. In order to circumvent this limitation, it is desirable to find novel schemes for generating optical lattices with spacing $\deff\ll\lambda$, in order to enhance the associated energy scale $\Ereff=h^2/(8m\deff^2)$ \cite{yi2008state}. Schemes have been proposed to generate lattices of sub-wavelength spacing, based on multi-photon optical transitions \cite{dubetsky2002lambda} or on adiabatic dressing of state-dependent optical lattices \cite{yi2008state}; the realization of  lattices with spacing $\deff=\lambda/4$  was reported in Ref.\,\cite{ritt2006fourier}. An interesting alternative would be to trap atomic gases in the electromagnetic fields of nano-structured condensed-matter  systems~\cite{gullans2012nanoplasmonic,thompson2013coupling,romero2013superconducting}.

\newcommand{\lenD}{1.1cm}
\newcommand{\lenE}{0.85\linewidth}

\begin{figure}[t!]
\includegraphics[width=\linewidth]{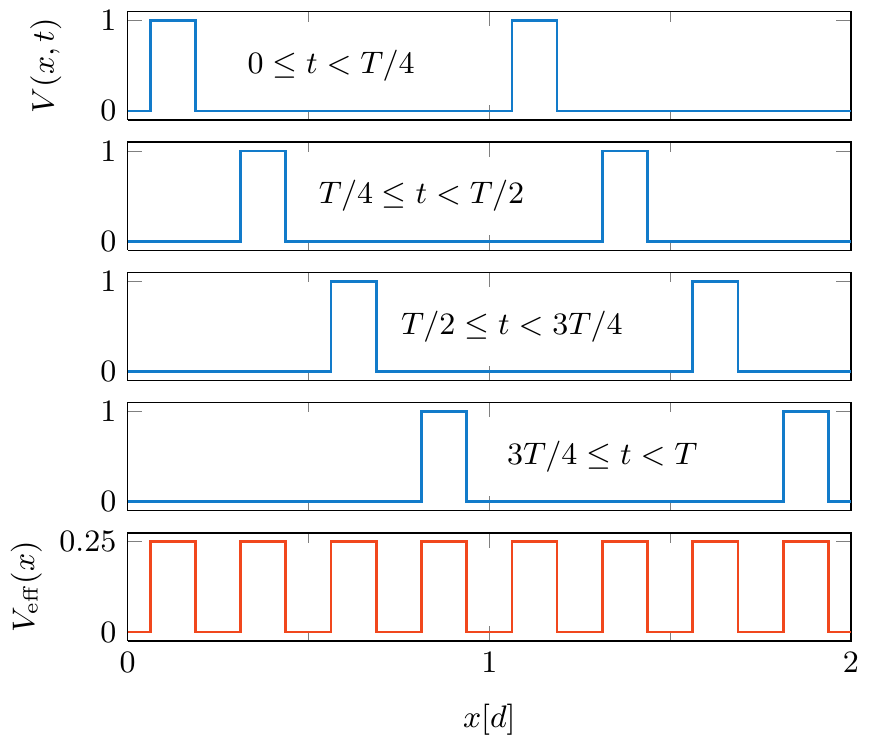}
\vspace{-6mm}
\caption{
Stroboscopic scheme for engineering short-spacing lattices, illustrated on the case $N=4$. We make use of a periodic potential $V(x,t)$ of spatial period $d$, that is shifted of the distance $d/N$ after every time step of duration $T/N$ (blue curves). The effective potential $V_{\mathrm{eff}}(x)$ (red curve), resulting from time averaging, exhibits a spatial period $\deff=d/N$. }
\label{Fig_scheme}
\end{figure}

In this letter, we propose a novel scheme leading to lattices of  spacing $\deff=d/N$, $N$ being an arbitrary integer, based on  spin-dependent lattices with time-periodic modulation. In the regime of large modulation frequency \cite{avan1976effect,rahav2003effective,goldman2014periodically,bukov2014universal}, the  atom dynamics is governed by an effective static periodic potential of spacing $\deff$, with an additional micro-motion. This description is confirmed by a numerical simulation, which shows the possibility to load adiabatically the ground state of the effective lattice and to perform Bloch oscillations. We discuss the extension of the scheme to  two-dimensional lattices with non-trivial topology. Lattices with artificial magnetic fields, generally leading to topological bands, were recently realized in experiments, with  standard lattice spacing \cite{goldman2014light}.
For those systems, increasing the energy scale using short-spacing lattices could prove important for creating strongly-correlated states such as fractional Chern insulators \cite{bergholtz2013topological,parameswaran2013fractional}.

A basic scheme of our method is pictured in Fig.\,\ref{Fig_scheme}. Consider a periodic potential $V(x)$ of period $d$, which is abruptly shifted by the distance $d/N$ at stroboscopic times $t_n=(n/N)T$, $n\in\mathbb{Z}$, leading to a time-periodic potential $V(x,t)$ of period $T$. Provided that $T$ is much smaller than typical timescales of atomic motion, the atoms experience an effective time-averaged potential $V_{\mathrm{eff}}(x)=\int V(x,t)\dd t/T$. A simple calculation shows that $V_{\mathrm{eff}}(x)$ is given by the sum of all harmonics of the potential $V(x)$, whose orders are multiples of $N$ \cite{SOM}\nocite{hauke2012non,goldman2015periodically,denschlag2002bose}. The effective potential $V_{\mathrm{eff}}(x)$ is thus spatially periodic, of spatial period $\deff=d/N$.

Conventional optical lattices present a spatial modulation proportional to the intensity pattern of interfering light waves, which exhibit spatial frequencies of at most twice the light momentum $k$.  Thus, applying the stroboscopic scheme in Fig.~\ref{Fig_scheme} to these potentials could not lead to effective lattices of period $\deff<\lambda/2$. This restriction does not apply to spinful particles subjected to spin-dependent optical lattices. As an illustration, consider a spin-$1/2$ particle evolving in the potential $V(x)=\VL\cos(2kx)\sigma_z+\VB\sigma_x$, where $\sigma_u$ ($u=x,y,z$) are the Pauli matrices. In a dressed state picture, the atom may follow adiabatically the state of lowest energy $V_-(x)=-\sqrt{\VL^2\cos^2(2kx)+\VB^2}$.
 As this potential  exhibits harmonics of the spatial frequency $2k$ of all orders, the lattice spacings achievable by applying the stroboscopic scheme to $V_-(x)$ can be made arbitrarily small.

We describe in the following a modified, more practical, version of this scheme, which consists of a spin-dependent optical lattice with smooth temporal variations, given by
\begin{equation}\label{eq_potential}
V(x,t)=\VL\cos(2kx-\Omega t)\sigma_z+\VB\cos(N\Omega t)\sigma_x.
\end{equation}
This potential satisfies $V(x+d/N,t+T/N)=V(x,t)$, with $d=\pi/k$, thus, it can be viewed as a continuous version of the stroboscopic scheme.  
Understanding the physical effects of the potential (\ref{eq_potential}) falls within the description of time-periodic Hamiltonian systems \cite{avan1976effect,rahav2003effective,goldman2014periodically,bukov2014universal}. Following Ref.\,\cite{goldman2014periodically}, we describe the dynamics of an atom between the times $\ti$ and $\tf$ as
 \begin{equation}\label{eq_U}
U(\ti\rightarrow \tf)=e^{-iK(\tf)}e^{-\frac{i}{\hbar}(\tf-\ti)\Heff}e^{iK(\ti)},
\end{equation}
where we introduce a time-independent, effective Hamiltonian $\Heff$ and a time-periodic kick operator $K(t)$. The three operators in (\ref{eq_U}) describe, from right to left, the role of the initial phase of the Hamiltonian at time $\ti$, the evolution from $\ti$ to $\tf$ according to a stationary Hamiltonian, and the micro-motion related to the final phase of the Hamiltonian at time $\tf$.

The expressions for the effective Hamiltonian $\Heff$ and kick operator $K(t)$ can be calculated  through a perturbative expansion in powers of $1/\Omega$, see Refs.~\cite{rahav2003effective,goldman2014periodically}. To lowest-order, this yields
\rightline{%
\begin{minipage}[c]{1.017\linewidth}
\begin{align}
\Heff&=\frac{p^2}{2m}+\Veff(x),\label{eq_Heff}\\
\Veff(x)&=\frac{\Ueff}{2}\cos(2Nkx)\sigma_x,\quad\Ueff=\frac{2\VB}{N!}\left(\frac{\VL}{\hbar\Omega}\right)^N,\label{eq_Veff}\\
K(t)&=\frac{-\VL}{\hbar\Omega}\sin(2kx-\Omega t)\sigma_z+\frac{\VB}{N\hbar\Omega}\sin(N\Omega t)\sigma_x.\label{eq_K}
\end{align}
\smallskip
\end{minipage}
}
The effective potential (\ref{eq_Veff}), which describes a periodic potential of depth $\Ueff$ and spatial period $\deff=d/N$, was derived under the assumption that $N$ is an even integer (a similar potential is found for $N$ odd). The expression (4) has been obtained based on a Born-Oppenheimer approximation, in which the kinetic energy term is neglected and one calculates the effective potential for a given position $x$, treating internal degrees of freedom ($\sigma_j$ operators) quantum-mechanically.
One finds that the terms neglected here are smaller than those given in (\ref{eq_Veff}), by a factor $\Er/(\hbar\Omega)\ll 1$.  Furthermore, this approximation is validated by a direct comparison with the full quantum treatment (see below).
In the Supplementary material we show that the perturbative expansion can be resumed, with respect to either the variable $\VL/(\hbar\Omega)$ or $\VB/(\hbar\Omega)$. There we also discuss the generalization of this modulation scheme to an arbitrary spin $F$, through the substitution  $\sigma_u \rightarrow 2 F_u$ \cite{SOM}.


\newlength{\lenA}
\setlength{\lenA}{9.8mm}

\begin{figure}
\includegraphics[width=1.02\linewidth]{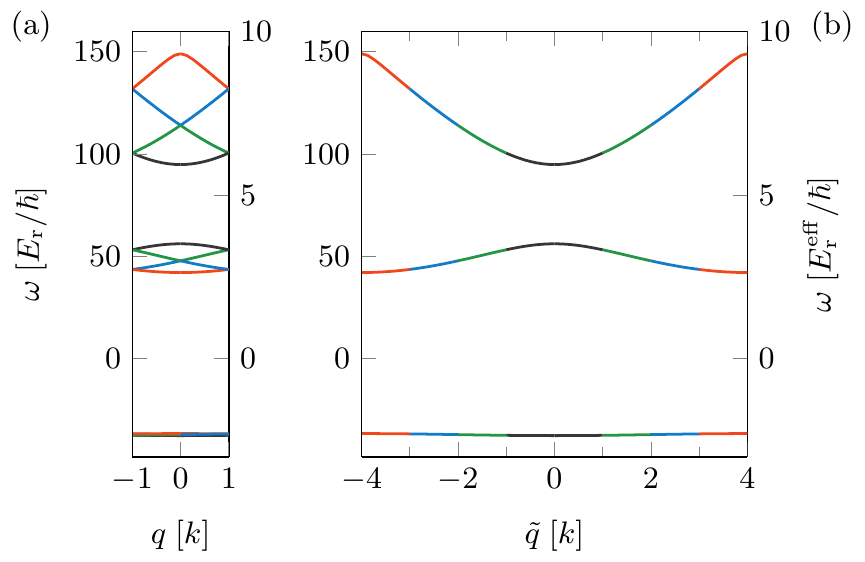}
\vspace{-6mm}
\caption{\label{Fig_BandStructure}Band structure of a  dynamic optical lattice of spacing $\deff=d/4$, corresponding to the parameters $N=4$, and $\VL=\VB=\hbar\Omega=200\,\Ereff$. In $(a)$, we make use of the spatial and temporal  translational symmetries $\mathcal{T}_x$, $\mathcal{T}_t$ and label the eigenstates by their quasi-momentum $-k\leq q<k$ and quasi-energy  $- \hbar \Omega/2\leq\hbar \omega<\hbar \Omega/2$.  The Bloch-Floquet bands can be unfolded using the additional symmetry $\mathcal{T}^*$, leading to the band structure in (b), indexed by the  modified quasi-momentum $-4k\leq \tilde{q}<4k$. The unfolding of the band structure can be followed from the different coloring of successive bands.}
\end{figure}

In order to test the validity of the effective Hamiltonian (\ref{eq_Heff}), we performed a numerical study of the full time-periodic Hamiltonian using the Floquet formalism. Since the Hamiltonian $H$ is invariant under the symmetries $\mathcal{T}_x:x\rightarrow x+d$ and $\mathcal{T}_t:t\rightarrow t+T$, we look for eigenstates written as Bloch-Floquet wave functions $\psi_{q,\omega}(x,t)=e^{i(qx-\omega t)}u_{q,\omega}(x,t)$, where $u_{q,\omega}(x,t)$ is $d$-periodic in $x$ and $T$-periodic in $t$ \cite{holthaus1992quantum,grifoni1998driven}. Eigenstates are labelled by their quasi-momentum $-k<q\leq k$ and quasi-energy  $0\leq \hbar \omega<\hbar \Omega$. An example of the band structure calculated numerically for $N=4$ is plotted in Fig.\,\ref{Fig_BandStructure}a. The band structure exhibits gap openings once every four bands, at the momenta $Nkp$, where $p\in\mathbb{Z}^*$,  as expected for a lattice of spacing $d/N$. 

The band structure can be unfolded, making use of the additional symmetry $\mathcal{T}^*: x\rightarrow x+d/N, t\rightarrow t+T/N$. As explained in the Supplementary Material, eigenstates  associated with the symmetries $\mathcal{T}_x$, $\mathcal{T}_t$ and $\mathcal{T}^*$ can be written as $\psi_{\tilde{q},\omega}(x,t)=e^{i(\tilde{q}x-\omega t)}v_{\tilde{q},\omega}(x,t)$, where $v_{\tilde{q},\omega}(x,t)$ is $d/N$-periodic in $x$ and $2\pi$-periodic in $(kx-\Omega t)$  \cite{SOM}. We show  the band structure calculated within this formalism in Fig.~\ref{Fig_BandStructure}b, which is very close to that expected for a  lattice of spacing $d/N$ and depth $\Ueff\simeq10.9\,\Ereff$ 
\footnote{This depth value slightly differs from the perturbative result  $\Ueff\simeq16.7\,\Ereff$, from eq.\,(\ref{eq_Veff}), since $V_{L,B}\not\ll \hbar\Omega$. We checked numerically that the difference can be accounted for by  higher-order terms.}
.

\begin{figure}
\includegraphics[width=1.02\linewidth]{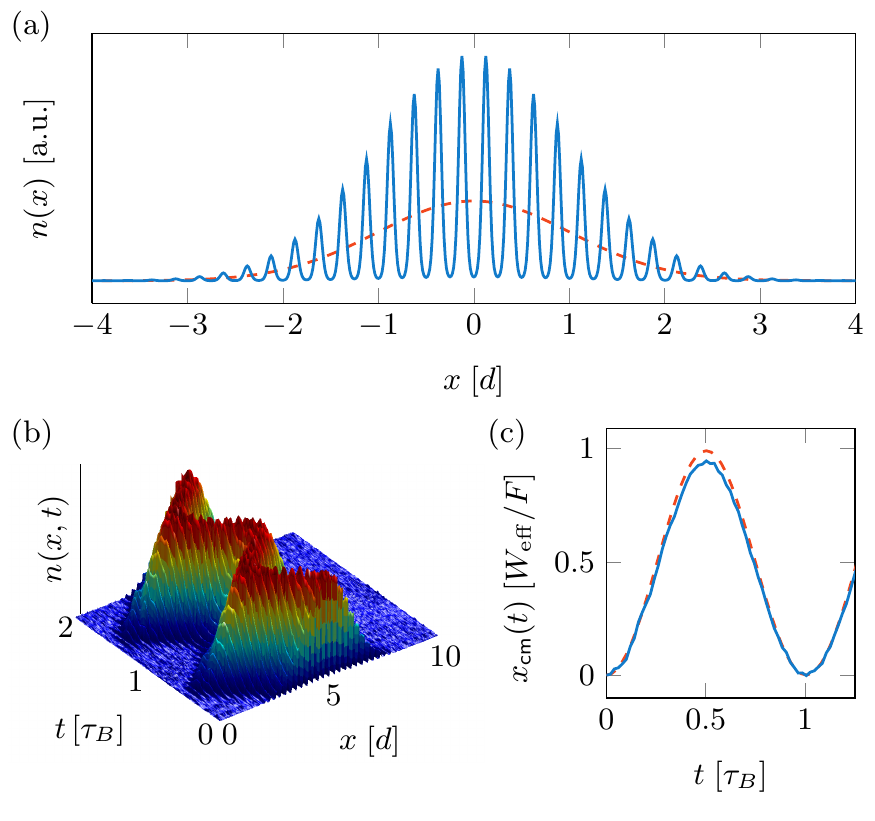}
\vspace{-8mm}
\caption{(a) Atomic density of a  wave-packet loaded into a dynamic lattice of spacing $\deff=d/4$. We start from a gaussian  wave-packet, spin-polarized along $x$, of wave function  $\psi(x,t\!=\!0)=\exp[-x^2/(2\sigma^2)]$, with  $\sigma\simeq1.4\,d$ (red dashed line). The lattice depth $\VL$ is slowly  ramped up for a duration $\tramp=20\hbar/\Ereff$ from $\VL=0$ to $\VL=\VL^0$, and  lattice parameters  $N=4$, $\VL^0=\VB=\hbar\Omega=200\,\Ereff$. The atom density after loading is spatially modulated, with a period $d/4$ (blue line). (b) Evolution of the density distribution during Bloch oscillations, calculated for the dynamic lattice parameters of (a), and for a force  $F=W_{\text{eff}}/(8\,\deff)$, where  $W_{\text{eff}}\simeq 0.06\,\Ereff$ is the expected bandwidth of the lowest band for $U_{\text{eff}}=10.9\,\Ereff$. (c) Evolution of the center-of-mass position during Bloch oscillations, calculated for a standard optical lattice of depth  $U_{\text{eff}}=10.9\,\Ereff$ (red dashed line), and for the dynamic optical lattice (blue line). The time and spatial coordinates are plotted in units of the ideal Bloch period $\tau_B=2N\hbar k/F$ and amplitude  $W_{\text{eff}}/F$.}
\label{FigBlochOsc}
\end{figure}

The practical relevance of the short-spacing lattice described above is based on the ability to load atoms into the ground state of the effective potential  (\ref{eq_Veff}). 
 The analysis of this loading protocol requires special care, as the effective-Hamiltonian approach inherent to Eq.~(\ref{eq_U}) assumes a constant lattice depth~\cite{goldman2014periodically}. In fact, we find that the concept of the effective Hamiltonian can be modified so as to describe the time-evolution under a ramp of the moving-lattice depth $\VL$, see  Ref.~\cite{SOM}. We simulate the lattice  loading from a numerical calculation of the full dynamics of an atomic wave packet under the action of the potential $(\ref{eq_potential})$. 
Starting from a gaussian wave packet, spin-polarized along $x$, we solve  the Schr\"odinger equation, discretized in space and time, with a lattice depth $\VL$ slowly ramped up for a duration $\tramp$. 
 As shown in Fig.\,\ref{FigBlochOsc}a, a ramp duration $\tramp=20\hbar/\Ereff$ leads to a state with strong spatial modulations of spacing $d/N$, as expected for a wavepacket prepared in the lowest band of the effective lattice (\ref{eq_Veff}). The calculated population in the effective lowest band is 93\,\%, close to the value expected with standard optical lattices for such a ramp duration. In the Supplementary Information we analyze the momentum distribution, which corresponds to the one expected for the ground state of the effective lattice, slightly modified by the micro-motion  \cite{SOM}. 

The system description as an effective $d/N$ lattice is also supported by a numerical simulation of Bloch oscillations. We calculate the action of a linear potential $-Fx$ applied to the state obtained after the lattice loading. As shown in Fig.\,\ref{FigBlochOsc}b,c, the wave packet undergoes Bloch oscillations, revealed as real-space oscillations of its center of mass. Both the amplitude and period of this oscillation agree well with those expected for an effective lattice of period $d/N$ and depth  $\Ueff$ inferred from band structure calculations.

\newlength{\lenB}
\setlength{\lenB}{46.5mm}
\newlength{\lenC}
\setlength{\lenC}{0.3pt}
\begin{figure}
\includegraphics[height=0.8\linewidth]{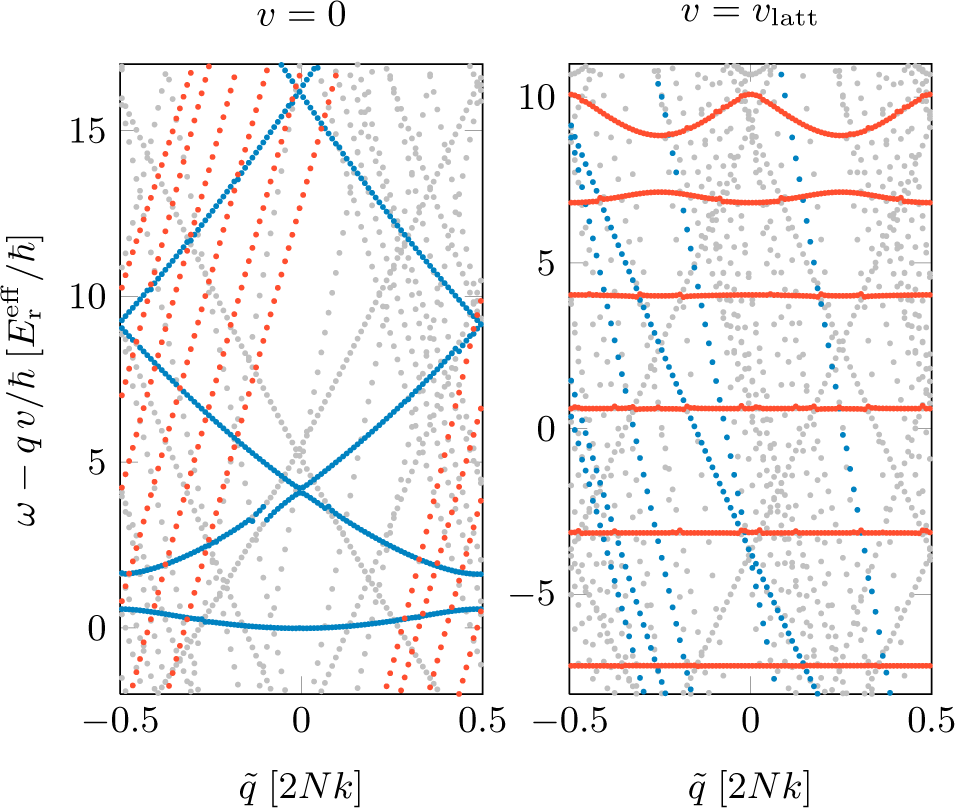}
\vspace{-2mm}
\caption{Band structure corresponding to the dynamic lattice for the parameters $N=2$, $\VL=\VB=\hbar\Omega=10\,\Ereff$. The  panels correspond to different frames of reference, of velocity $v=0$ (left) and $v=\vlatt=\Omega/(2k)$ (right). The blue points correspond to the band structure of an optical lattice of spacing $\deff=d/N$ and depth $\Ueff\simeq2\,\Ereff$, at rest in the laboratory frame. The red dots correspond to the band structure of an optical lattice of spacing $d$ and depth $U\simeq74\,\Er$ ($\simeq9\,\Ueff$), at rest in the frame of velocity $v=\vlatt$ \cite{RenormU}.\label{Fig_ChangeFrame}}
\end{figure}

The potential $V(x,t)$ written in (\ref{eq_potential}) corresponds to the sum of a time-modulated magnetic field and a spin-dependent optical lattice moving at the velocity $\vlatt=\Omega/(2k)$. In the above discussion we considered the effect of this potential as an effective static optical lattice. An alternative view is obtained in the frame of reference moving at the velocity  $v=\vlatt$, where the potential $V(x'=x-vt,t)$ consists of the sum of a modulated magnetic field and a very deep static lattice $\VL\cos(2kx')\sigma_z$, with $\VL\sim\hbar\Omega\gg\Ueff$. Both points of view can be reconciled by a proper interpretation of the band structure, as illustrated for the case $N=2$ in  Fig.\,\ref{Fig_ChangeFrame}. Among the eigenenergies $\omega(\tilde{q})$ calculated numerically in the laboratory frame $v=0$, we identify the Bloch bands corresponding to a static effective lattice of spacing $\deff$. The eigenenergies $\omega'(\tilde{q})$ corresponding to a frame of reference moving at a velocity $v$ can be deduced from those in the laboratory frame using  the relation  $\omega'=\omega-\tilde{q}v/\hbar$. In the frame moving at  $v=\vlatt$, we observe Bloch bands corresponding to a  very deep static optical lattice of period $d$.

\newcommand{\radd}{0.08}

\begin{figure}[position]
\includegraphics[width=\linewidth]{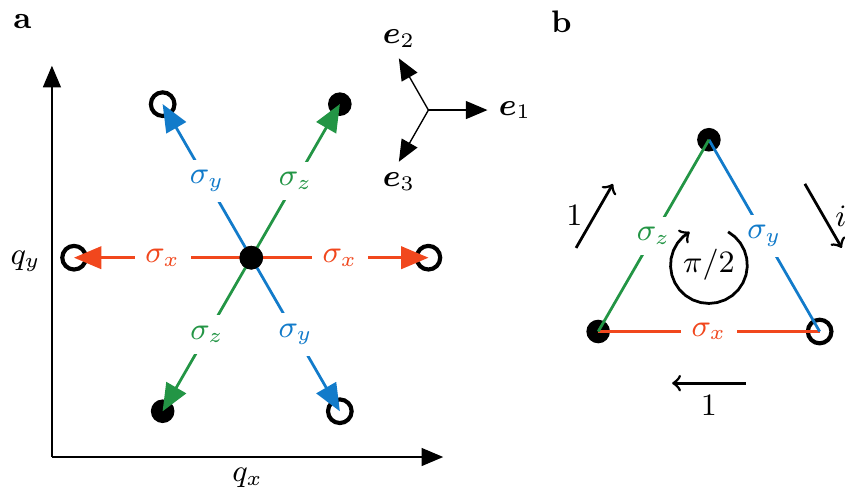}
\vspace{-5mm}
\caption{(a) Momentum-space representation of the effective couplings in Eq. \ref{eff_2D},  illustrated as arrows of length $2Nk$, oriented along the unit vectors $\pm\bs{e}_{i}$ ($i=1,2,3$), and proportional to Pauli matrices. Quantum states are represented in the basis $\{\ket{+_z}$ (filled dots), $\ket{-_z}$ (circles)$\}$.  (b) Phase accumulated around a
triangular subcell of the $k$-space lattice. Due to the internal-state degree of freedom, the
unit cell of the lattice is formed by four triangular subcells. The same phase of $\phi=\pi/2$ is
found to be accumulated around all subcells, indicating
that the lowest energy band is associated with a non-trivial Chern
number $\nu_{\text{Ch}}=1$ \cite{cooper2012designing}.
}
\label{fig1}
\end{figure}

We now consider a 2D extension of our scheme. The time-dependent part of the Hamiltonian is taken as
\begin{align}
V (\rr,t)&=\VL\cos (2k \bs{e}_1\cdot\rr - \Omega_1 t) \hat \sigma_z+ \VB \cos ( N \Omega_1 t) \hat \sigma_x \notag \\
&+\VL \cos (2k \bs{e}_2\cdot\rr - \Omega_2 t) \hat \sigma_x+ \VB\cos ( N \Omega_2 t) \hat \sigma_y \notag \\
&+\VL\cos (2k \bs{e}_3\cdot\rr - \Omega_3 t) \hat \sigma_y+ \VB \cos ( N \Omega_3 t) \hat \sigma_z, \notag \end{align}
where the unit vectors $\bs{e}_{1,2,3}$ have directions as represented in Fig. \ref{fig1}. For a suitable choice of the frequencies $\Omega_{1,2,3}$ \cite{ChoiceFreq}, each line of the equation above can be treated individually, which results in an effective potential of the form  \cite{cooper2011optical}
\begin{align}
\Veff(\rr)\simeq \frac{\Ueff}{2}[&\cos (2Nk \bs{e}_1\cdot\rr)\sigma_x +\cos (2Nk \bs{e}_2\cdot\rr)\sigma_y  \notag \\
&+\cos (2Nk \bs{e}_3\cdot\rr)\sigma_z], \label{eff_2D}
\end{align}
where $N$ is taken to be an even integer. These couplings are illustrated in quasi-momentum space in Fig.\,\ref{fig1}a. Following Ref.\,\cite{cooper2012designing}, the topological Chern number associated with the lowest energy band can be readily obtained from these couplings. Indeed, the Chern number measures the flux of the Berry curvature $\Omega (\bs q)$ over the entire (momentum-space) unit cell:
\be
\nu_{\text{Ch}}=\frac{1}{2\pi} \int_{\text{unit cell}} \Omega (\bs q) \text{d}^2 q,
\ee
which can be directly evaluated by calculating the phases accumulated by a state as it performs a loop around the triangular subcells \cite{cooper2012designing}.  For the effective lattice described by eq.\,\eqref{eff_2D}, each unit cell is constituted of four triangular subcells, and we find an accumulated phase of $\pi/2$ within each of them  (see Fig.\,\ref{fig1}b). In this configuration, the Chern number of the lowest band is given by $\nu_{\text{Ch}}=(1/2\pi)\times 4\times(\pi/2)=1$. Generally the reasoning above is valid only in the  weak-binding regime; however, for the coupling (\ref{eff_2D}), $\nu_{\text{Ch}}$ is unchanged for all values of $\Ueff$. Note that the size of the unit cell in real space scales as $1/N^2$; we thus expect the flux density to be increased by a factor of $N^2$ compared to standard optical lattices.

The method discussed above is based on applying strong spin-dependent optical lattices, for which Lanthanide atoms would be most suited for a practical implementation. Indeed, the optical lattices could be generated using laser light close to a narrow optical transition, which would lead to deep, spin-dependent lattices with negligible Rayleigh scattering effects (for Dy atoms, one can achieve ratios $\hbar\Gamma_{\mathrm{scattering}}/\VL\sim 10^{-7}$) \cite{nascimbene2013realizing,cui2013synthetic,dalibard2014introduction}.

In conclusion, we introduced a novel scheme to engineer spatially periodic atom traps of sub-wavelength spacing, based on the application of spin-dependent optical lattices. 
A natural extension of this work would be to include interactions between atoms in the effective lattice description, and to understand whether micro-motion plays a significant role in scattering properties \cite{choudhury2014stability,bilitewski2014scattering,eckardt2015consistent}. This aspect will play a central role for investigating quantum many-body physics with short-spacing lattices.

The authors are pleased to acknowledge Fabrice Gerbier and J\'er\^ome Beugnon
for valuable discussions. This work is supported by IFRAF, ANR (ANR-12-BLANAGAFON),
ERC (Synergy UQUAM), the Royal Society of London and the EPSRC. N.G. is financed by the FRS-FNRS Belgium and by the BSPO under the PAI project P7/18 DYGEST.


\begin{thebibliography}{36}
\bibitem [{Cho()}]{ChoiceFreq}%
  \BibitemOpen
  \href@noop {} {}\bibinfo {note} {The frequencies $\Omega_{1,2,3}$ should not
  be chosen too close to each other, and their ratios should not approach
  simple fractions.}\BibitemShut {Stop}%
\end{thebibliography}

%

\clearpage



\section*{Supplementary material}

# Supplementary Material for
# Dynamic optical lattices of sub-wavelength spacing for ultracold atoms

Sylvain Nascimbene,[1,][*] Nathan Goldman,[1, 2] Nigel Cooper,[3] and Jean Dalibard[1]

[1]*Laboratoire Kastler Brossel, Collège de France, ENS-PSL Research University,*
*CNRS, UPMC-Sorbonne Universités, 11 place Marcelin Berthelot, 75005 Paris, France*
[2]*CENOLI, Faculte des Sciences, Université Libre de Bruxelles (U.L.B.), B-1050 Brussels, Belgium*
[3]*T.C.M. Group, Cavendish Laboratory, J.J. Thomson Avenue, Cambridge CB3 0HE, United Kingdom*

(Dated: October 2, 2015)

## S.I. EFFECTIVE POTENTIAL CREATED BY THE STROBOSCOPIC SCHEME

In this Section, we present the calculation of the effective lattice potential produced via the stroboscopic scheme illustrated in Fig. 1 (main text). We start from a periodic potential $V(x)$ of period $d$, which can be decomposed in Fourier series as

$$V(x) = \sum_{p\in\mathbb{Z}} V_p\, e^{i2\pi p x/d}.$$

The method consists in shifting the potential $V(x)$ of the distance $d/N$, after each time interval $T/N$. For $N$ integer, this leads to a time-periodic potential $V(x,t)$ of time period $T$. For a sufficiently short period $T$, the atomic motion is governed by the effective potential $V_{\text{eff}}(x)$, equal to the time average of $V(x,t)$:

$$\begin{aligned}
V_{\text{eff}}(x) &= \frac{1}{T}\int_0^T V(x,t)\mathrm{d}t \\
&= \frac{1}{N}\sum_{j=1}^N V(x + jd/N) \\
&= \sum_{p\in\mathbb{Z}} V_p\, e^{i2\pi px/d}\frac{1}{N}\sum_{j=1}^N e^{i2\pi pj/N} \\
&= \sum_{p\text{ multiple of }N} V_p\, e^{i2\pi px/d}.
\end{aligned}$$

It is then apparent that the effective potential $V_{\text{eff}}(x)$ is periodic, of period $d_{\text{eff}} = d/N$.

## S.II. CALCULATION OF THE EFFECTIVE POTENTIAL

In this section we give details about the establishment of the effective Hamiltonian (eqs. (3), (4) in the main text). We first provide a simple description of the effective Hamiltonian in a Born-Oppenheimer approximation,

---

[*] sylvain.nascimbene@lkb.ens.fr

and discuss further the relevance of this approximation for typical lattice parameters.

The Born-Oppenheimer approximation consists in neglecting the kinetic energy for the calculation of the effective Hamiltonian. The position $x$ is considered as a fixed parameter, while internal degrees of freedom are treated quantum mechanically. Using the formalism of ref. [S1], we decompose the Hamiltonian in Fourier series as

$$\begin{aligned}
H(t) &= H_0 + \sum_{j=1}^{\infty} V^{(j)} e^{ij\Omega t} + V^{(-j)} e^{-ij\Omega t}, \\
V^{(1)} &= V^{(-1)\dagger} = V_{\text{L}} e^{-2ikx}\sigma_z/2, \\
V^{(N)} &= V^{(-N)\dagger} = V_{\text{B}}\sigma_x/2, \\
V^{(j)} &= 0 \quad \text{otherwise.}
\end{aligned}$$

The effective Hamiltonian can be expanded as a series in $1/\Omega$. At lowest order, it reads

$$\begin{aligned}
H_{\text{eff}} &= \frac{p^2}{2m} + V_{\text{eff}}(x), \\
V_{\text{eff}}(x) &= \frac{1}{N!(\hbar\Omega)^N}\left[V^{(1)},\ldots,\left[V^{(1)},V^{(-N)}\right]\right] + \text{h.c.},
\end{aligned}$$

with $V^{(1)}$ occurring $N$ times. One calculates

$$\begin{aligned}
V_{\text{eff}}(x) &= \frac{V_{\text{L}}^N V_{\text{B}} e^{-i2Nkx}}{2N!(2\hbar\Omega)^N}[\sigma_z,\ldots,[\sigma_z,\sigma_x]] + \text{h.c.} \\
&= \frac{U_{\text{eff}}}{2}\begin{cases} \cos(2Nkx)\sigma_x, & N \text{ even,} \\ \sin(2Nkx)\sigma_y, & N \text{ odd,} \end{cases}
\end{aligned}$$

where $U_{\text{eff}} = 2V_{\text{B}}(V_{\text{L}}/\hbar\Omega)^N/N!$.

In the main text we consider the case of a (pseudo)-spin-1/2 for simplicity. The scheme can directly be extended to an arbitrary spin $F$. Assuming a coupling $V(x,t) = 2V_{\text{L}}\cos(2kx - \Omega t)F_z + 2V_{\text{B}}\cos(N\Omega t)F_x$, we make use of the general commutation algebra of $F_u$ operators and obtain an effective potential $V_{\text{eff}}(x) = U_{\text{eff}}\cos(2Nkx)F_x$ ($N$ even).

We now consider the validity of the Born-Oppenheimer approximation, which consists in neglecting the non commutation of position and momentum. The latter plays a role at the order $N + 1$ of the perturbative expansion

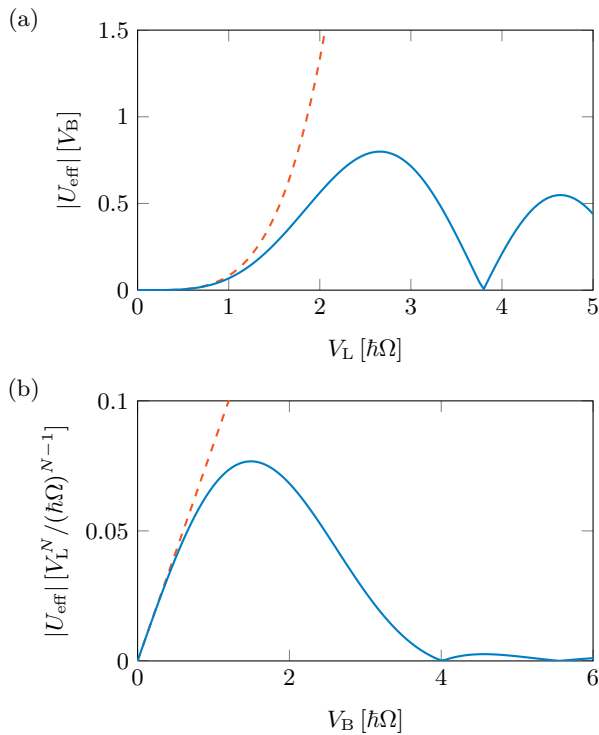

FIG. S1. Depth $U_{\text{eff}}$ of the effective lattice (in absolute value), calculated for arbitrary values of $V_L/(\hbar\Omega)$ (a) or $V_B/(\hbar\Omega)$ (b) in the case $N = 4$. The dashed lines correspond to the lowest-order perturbation result (4), and the solid lines to the resummation results (S.1) and (S.2).

in $1/\Omega$. We performed the perturbative expansion up to the order $N+1$ for the case $N = 1$, and obtained the expression

$$H_{\text{eff}} = \frac{p^2}{2m} + \frac{U_{\text{eff}}}{2}\sin(2kx)\sigma_y$$
$$+ \left(\frac{V_L}{\hbar\Omega}\right)^2 E_r + \frac{U_{\text{eff}}}{2}\frac{\hbar k}{2m\Omega}\left(p\sin(2kx) + \sin(2kx)p\right)\sigma_y.$$

While the first term of order $N+1$ only represents an energy offset, the second term can be viewed as a spin-orbit coupling which may affect the atom dynamics. However, for typical momenta $p$ on the order of the lattice momentum $k$, the amplitude of this term is smaller than the effective lattice depth $U_{\text{eff}}$ by a factor $\sim E_r/(\hbar\Omega)$, which is typically very small for the examples considered in this Letter.

## S.III. RESUMMATION OF THE PERTURBATIVE EXPANSION OF $H_{\text{eff}}$

The effective potential $V_{\text{eff}}$ can be calculated as a series expansion in powers of the (potentially small) dimensionless parameters $V_L/(\hbar\Omega)$ and $V_B/(\hbar\Omega)$, in the high-frequency limit $\Omega \to \infty$. In the main text, we provide its expression in Eq. (4), which corresponds to the lowest-order term. We note that this derivation, which is based on the general formula of Ref. [S1], was obtained by neglecting the non-commutativity of the position and momentum operators; indeed, we verified that the momentum operator is irrelevant in the derivation of the effective potential, which essentially relies on the spin-dependent time-modulated components of the Hamiltonian. Thus, in the following of this Section, which aims to derive the effective potential in the strong-modulation regime, we explicitly neglect any effects associated with the kinetic energy term of the full Hamiltonian.

In this Section, we first derive the expression for the effective potential $V_{\text{eff}}$, in the case where $V_L/(\hbar\Omega)$ is allowed to take arbitrary large values (still assuming that $V_B \ll V_L, \hbar\Omega$). Following Refs. [S2, S3], we perform a unitary transformation

$$|\psi'\rangle = R(t)|\psi\rangle, \quad R(t) = \exp\left(-i\frac{V_L}{\hbar\Omega}\sin(kx - \Omega t)\sigma_z\right),$$

which removes the diverging term $\sim V_L \sim \hbar\Omega$ from the time-dependent potential $V(x,t)$ in Eq. 1 (main text). This leads to a novel time-dependent potential

$$V'(x,t) = R(t)V(x,t)(t)R^{\dagger}(t) + i\hbar\partial_t R(t)R^{\dagger}(t)$$
$$= R(t)\left[V_B\cos(N\Omega t)\sigma_x\right]R^{\dagger}(t).$$

Making use of the identity $e^{-i\gamma\sigma_z}\sigma_x e^{i\gamma\sigma_z} = \cos(2\gamma)\sigma_x + \sin(2\gamma)\sigma_y$, we obtain the expression

$$V'(x,t) = V_B\cos(N\Omega t)\left[\cos\left(\frac{2V_L}{\hbar\Omega}\sin(kx - \Omega t)\right)\sigma_x\right.$$
$$\left.+ \sin\left(\frac{2V_L}{\hbar\Omega}\sin(kx - \Omega t)\right)\sigma_y\right].$$

In the large-frequency limit $\Omega \to \infty$, the atom dynamics can be described by an effective stationary potential, given by [S2, S3]

$$V_{\text{eff}}(x) = \frac{1}{T}\int_0^T V'(x,t)dt$$
$$= J_N\left(\frac{2V_L}{\hbar\Omega}\right)V_B\cos(2Nkx)\sigma_x, \quad (\text{S.1})$$

assuming $N$ even, and where $J_N$ is a Bessel function of the first kind. This effective potential corresponds to a spin-dependent optical lattice of spacing $d/N$ and depth $U_{\text{eff}} = 2J_N\left(\frac{2V_L}{\hbar\Omega}\right)V_B$.

A similar resummation with respect to $V_B/(\hbar\Omega)$ can also be derived. Here, we make use of the Floquet representation of time-periodic Hamiltonians. We first write the exact eigenstates of the coupling $\frac{V_B}{2}\cos(N\Omega t)\sigma_x$, which read

$$||n,s_x\rangle = \sum_{p \in \mathbb{Z}} J_p\left(\frac{2s_x V_B}{\hbar\Omega}\right)|n+pN,s_x\rangle,$$

where $n$ denotes the Floquet quantum number, and $s_x$ is the spin projection along $x$. The energy of the state $||n, s_x\rangle$ is equal to $n\hbar\Omega$. The effect of the coupling $V_L \cos(kx - \Omega t)\sigma_x$ can be understood using perturbation theory in the degenerate subspace $||n, \pm\rangle$, which must be performed at order $N$. We obtain the expression

$$V_{\text{eff}}(x) = \frac{U_{\text{eff}}}{2} \cos(2Nkx)\sigma_x,$$

$$U_{\text{eff}} = 4\hbar\Omega \left(\frac{V_L}{2\hbar\Omega}\right)^N \left| \sum_{\sum_{i=1}^N p_i = -1} \frac{\prod_{i=1}^N J_{p_i}[(-1)^i \frac{2V_B}{N\hbar\Omega}]}{\prod_{i=1}^{N-1} \sum_{j=1}^i (1 + Np_j)} \right|.$$
(S.2)

We plot in Fig. S1 the lattice depth $U_{\text{eff}}$ given by the re-summation formulas in Eqs. (S.1)-(S.2) discussed above. We checked that the formulas (S.1) and (S.2) account well for the numerical results obtained via direct diagonalization of the Bloch-Floquet equations (see Section S.IV).

## S.IV. EXPRESSION FOR THE BLOCH-FLOQUET HAMILTONIAN

The modulated potential (1) is invariant under the space and time translational symmetries $\mathcal{T}_x$, $\mathcal{T}_t$ and $\mathcal{T}^*$, which all commute with each other. The eigenstates of the Hamiltonian can thus be written as eigenstates of those symmetries, which can be expressed as

$$\psi_{\tilde{q},\omega}(x,t) = e^{i(\tilde{q}x - \omega t)} \sum_{j,l \in \mathbb{Z}} c_{j,l} \, e^{il(kx - \Omega t)} e^{ijNkx},$$

where $-Nk < \tilde{q} \leq Nk$ and $0 \leq \omega < \Omega$. The spinor coefficients $c_{j,l}$ are determined by the equations

$$\hbar(\omega + l\Omega)c_{j,l} = \frac{\hbar^2[\tilde{q} + (l + Nj)k]^2}{2m} c_{j,l}$$
$$+ \frac{V_L}{2}\sigma_x(c_{j,l+1} + c_{j,l-1})$$
$$+ \frac{V_B}{2}\sigma_z(c_{j+1,l-N} + c_{j-1,l+N}).$$

The numerical data represented in Fig. 2 (main text) is calculated using the above equations, in a truncated basis $-10 \leq j, l \leq 10$.

## S.V. MICRO-MOTION EFFECTS IN THE MOMENTUM DISTRIBUTION

In this section, we analyze how the micro-motion associated with the time-modulation in Eq. (1) affects the momentum distribution of atoms prepared in the effective potential $V_{\text{eff}}$ of spatial period $d/N$ [Eq. (4)]. Specifically, we consider an atom prepared in the ground state of the effective potential. This state can be expanded on

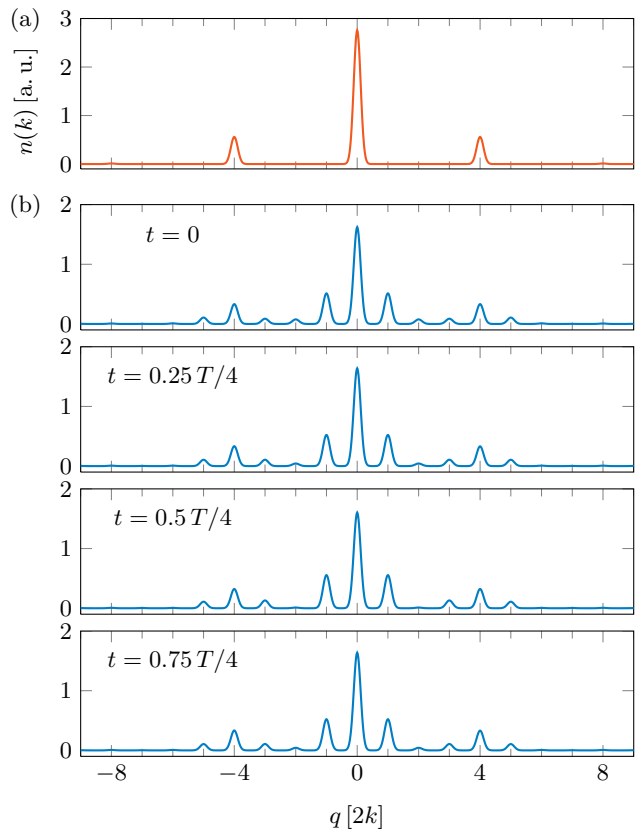

FIG. S2. (a) Momentum distribution associated with the ground state of the effective lattice with spacing $d/4$ and depth $U_{\text{eff}} = 10.9 \, E_r^{\text{eff}}$. (b) Momentum distribution of the state in (a), taking into account the micro-motion expected for the dynamic lattice parameters, according to Eq. S.3. The micro-motion leads to a more complex structure compared to (a), periodically evolving in time. The lattice parameters correspond to the ones of Fig. 2 in the main text.

the family of states of momentum multiple of $2Nk$ (see Fig. S2a).

The actual state created using time-modulated lattices is expected to be modified by the micro-motion, as

$$|\psi(t)\rangle = e^{-iK(t)} |\psi_0\rangle, \qquad (S.3)$$

where the expression for the kick operator $K(t)$ is given in the main text [Eq. (5)]. The latter leads to additional diffraction peaks at all momenta multiple of $2k$, whose amplitude vary periodically in time, with a period $T/N$ (see Fig. S2b). This shows that Bragg diffraction does not give a direct information on the ground state of the effective lattice.

## S.VI. EFFECTIVE HAMILTONIAN DURING LATTICE LOADING

In this Section, we analyze the adiabatic preparation of the ground state associated with the effective potential

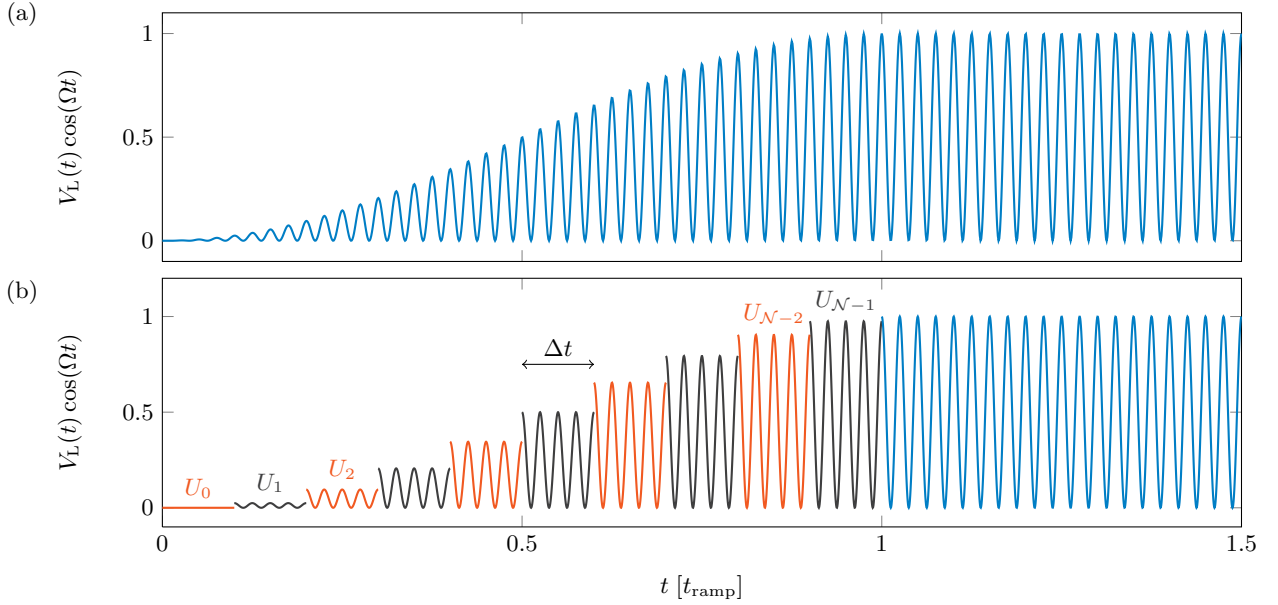

FIG. S3. (a) Evolution of the amplitude of the moving optical lattice at position $x=0$, during and after the lattice ramp of duration $t_\text{ramp}$. (b) Scheme of the ramp discretization: the time interval $0 \leq t \leq t_\text{ramp}$ is decomposed into $\mathcal{N}$ steps of duration $\Delta t$. Within each step the depth $V_\text{L}$ is constant, leading to a time-periodic potential.

$V_\text{eff}(x)$ of depth $U_\text{eff}^0$. We consider a slow ramp of the moving-lattice depth $V_\text{L}(t) = 0 \to V_\text{L}^0$ during the time interval $0 \leq t \leq t_\text{ramp}$, such that the effective potential's depth $U_\text{eff}^0$ corresponds to the final value $V_\text{L}(t_\text{ramp}) = V_\text{L}^0$. As the definition (2) of the effective Hamiltonian and kick operators assumes a constant lattice depth [S1], we expect these notions to be modified during the ramp. It is the aim of this Section to show how the adiabatic ramp can still be captured by an effective-Hamiltonian picture.

To analyze this situation, we decompose the ramp into $\mathcal{N}$ steps, and we assume that the time interval $\Delta t = t_\text{ramp}/\mathcal{N}$ is short enough, such that $V_\text{L}$ can be considered to remain constant within each step. More precisely, we assume that the lattice depth is equal to $V_\text{L}(j\Delta t)$ during the step $j\Delta t \leq t < (j+1)\Delta t$. We then apply the effective-Hamiltonian formalism of Ref. [S1] within each time-step, and write the full time-evolution operator as

$$U_\text{ramp} = \prod_{j=\mathcal{N}-1}^{0} U_j,$$
$$U_j = e^{-iK_0[V_\text{L}(j\Delta t)]} e^{-iH_\text{eff}[V_\text{L}(j\Delta t)]\Delta t/\hbar} e^{iK_0[V_\text{L}(j\Delta t)]}.$$

In the latter expression, and for the sake of simplicity, we assumed that $\Delta t$ was a multiple of the modulation period, so that the kick operators at the beginning and at the end of each step only depend on the value of $V_\text{L}$ (in fact, they correspond to the kick operator at the time $t=0$, hence the notation $K_0$).

Assuming $\Delta t$ short enough, we write

$$e^{iK_0[V_\text{L}((j+1)\Delta t)]} e^{-iK_0[V_\text{L}(j\Delta t)]} \simeq e^{i\Delta t (\mathrm{d}V_\text{L}/\mathrm{d}t)\mathrm{d}K_0/\mathrm{d}V_\text{L}},$$

leading to

$$U_\text{ramp} = e^{-iK(t_\text{ramp})} \mathcal{T}\left\{\exp\left(-i \int H_\text{eff}^\text{ramp}(t)\mathrm{d}t/\hbar\right)\right\},$$

where $\mathcal{T}$ denotes time-ordering, and where one introduced the slowly varying Hamiltonian

$$H_\text{eff}^\text{ramp}(t) = H_\text{eff}|_{V_\text{L}(t)} - \hbar \frac{\mathrm{d}V_\text{L}}{\mathrm{d}t} \left.\frac{\mathrm{d}K_0(t_\text{ramp})}{\mathrm{d}V_\text{L}}\right|_{V_\text{L}(t)}$$
$$= \frac{U_\text{eff}(t)}{2}\cos(2Nkx)\sigma_x + \frac{1}{\Omega}\frac{\mathrm{d}V_\text{L}}{\mathrm{d}t}\sin(2kx)\sigma_z. \quad (\text{S.4})$$

We now estimate a criterion for identifying the adiabatic regime of the lattice loading. Describing the lattice loading solely with the first term of eq. (S.4) would lead to the standard adiabaticity criterion $t_\text{ramp} \gg \hbar U_\text{eff}^0/(E_\text{r}^\text{eff})^2$ [S4]. We expect the second term in (S.4) to drive non-adiabatic transitions for $\dot{V}_\text{L} \gtrsim \Omega E_\text{r}$. Adiabatic lattice loading thus requires the additional constraint $t_\text{ramp} \gg (V_\text{L}^0/E_\text{r})/\Omega$. As we choose $V_\text{L} \lesssim \hbar\Omega$, this constraint should not be the most restrictive.



\end{document}